# A Survey of Static Formal Methods for Building Dependable Industrial Automation Systems

Roopak Sinha[1], Sandeep Patil[2], Luis Gomes[3] and Valeriy Vyatkin[4]

[1]Department of IT and Software Engineering, Auckland University of Technology, Auckland, New Zealand

[2]Department of Computer Science, Electrical and Space Engineering
Luleå University of Technology, Luleå, Sweden

[3]Electrical and Computer Engineering Department, NOVA School of Science and Technology
NOVA University of Lisbon, Portugal, & Centre of Technology and Systems,
UNINOVA, Caparica 2829-517, Portugal

[4]Department of Computer Science, Electrical and Space Engineering, Luleå University of Technology,
Luleå 97187, Sweden, the Computer Technologies International Laboratory, ITMO University, Saint
Petersburg 197101, Russia, & the Department of Electrical Engineering in Automation, Aalto University,
Aalto 00076, Finland

emails: [1]roopak.sinha@aut.ac.nz, [2]sandeep.patil@ltu.se, [3]lugo@uninova.pt, [4]vyatkin@ieee.org

**Abstract**

*Industrial automation systems (IAS) need to be highly dependable; they should not merely function as expected but also do so in a reliable, safe, and secure manner. Formal methods are mathematical techniques that can greatly aid in developing dependable systems and can be used across all phases of the system development life cycle (SDLC), including requirements engineering, system design and implementation, verification and validation (testing), maintenance, and even documentation. This state-of-the-art survey reports existing formal approaches for creating more dependable IAS, focusing on static formal methods that are used before a system is completely implemented. We categorize surveyed works based on the phases of the SDLC, allowing us to identify research gaps and promising future directions for each phase.*

## 1. INTRODUCTION

Industrial automation systems (IAS) are highly distributed systems containing software to control mechatronic components interacting with physical processes. IAS find use in production, logistics, and energy generation and distribution. IAS need to provide a high level of *dependability*, which is defined as the trustworthiness of a computer system's ability to reliably provide the service it promises to deliver [1]. A dependable system must be *functionally* correct and do what it promises, but it must also meet *nonfunctional* requirements (NFRs), such as reliability, safety, and security. As system sizes and complexity grow, the lack of systematic methods can significantly limit our ability to build highly dependable systems. *Formal methods* include formalisms, algorithms, and processes that have sound mathematical foundations and, therefore, provide more objective and unambiguous means to model and check the dependability of a given system. Due to factors such as difficulty in their use, expert skills, significant manual effort required, and scalability concerns, only a few formal methods have found industrial use.

This paper surveys formal methods that can be used for dependability analysis of IAS. We restrict the scope of this study to only *offline* or static approaches for dependability [2], leaving out *online* or runtime approaches used for monitoring, diagnosis, and fault tolerance. A static approach abstracts a system or a part of a system specification into a model and then checks if the abstracted model is consistent with some notion of correctness. Such analysis can be automated, such as in model checking, or require human input, such as in guided theorem proving. An online approach observes the behavior of a system or a part of a system *during execution* while simultaneously checking if the execution so far is consistent with some notion of correctness. Static approaches tend to be more useful *before* a system is deployed, such as during system design and refinement. Runtime approaches can be used only when parts of the system are operational. The cost of finding and fixing issues in a system grows exponentially as its development progresses, which indicates that early use of static approaches can significantly reduce development time and costs. However, not all analyses can be completed before deployment, which makes the use of online approaches equally important. This paper focuses on static approaches because the body of knowledge covering both offline and online approaches is too expansive, and online approaches have been generally well studied elsewhere, such as in state-of-the-art surveys for fault diagnosis [3], [4], and event scheduling [5]. We, therefore, restrict this paper to surveying formal and static *fault prevention* and *fault removal* techniques [1]. Furthermore, we do not survey a few subtopics that have been covered in other surveys. These topics are controller synthesis [6] and formal methods for addressing networking issues [7], [8]. Existing surveys [9]–[10][11][12][13][14] are also loosely related to this survey, but focus narrowly on specific standards or IAS subdomains.

There are a number of well-established and widely understood challenges in assuring the dependability of IAS. *Scale* and *complexity* are the most obvious ones; typical IAS contain highly distributed and modularized software running on multiple programmable logic controllers (PLCs), and hundreds or even thousands of mechatronic components that must interact with often heterogeneous physical processes. Several concerns related to dependability exist throughout the system development life cycle (SDLC) for IAS. *Requirements engineering* involves eliciting and organizing requirements for a system, and dependability-related concerns in this phase include ensuring requirements are consistent, correct, and complete, and are managed efficiently throughout the subsequent SDLC phases. During *design*, dependability rests squarely on system architecture and hence key concerns include architecture selection, comparing design alternatives, and choosing a sound and scalable primary separation strategy. During design and *implementation* phases, ensuring consistency between subsequent refinements of a system is a key dependability-related concern. The *testing* phase demands scalable, comprehensive, and easy-to-use verification and validation techniques to test a system's dependability. Runtime management and reconfiguration of IAS are key dependability-related concerns in the final deployment stage. Throughout the SDLC, a key challenge is systematically complying with dependability standards, such as IEC 61508 for functional safety, by exploiting the structures of design and implementation standards, such as IEC 61499 or IEC 61131-3.

This survey reports a systematic mapping study (SMS) [15], as opposed to a systematic literature review, to identify the main areas of activity within this wide research topic. Following the SMS methodology, an initial search of peer-reviewed research covering individual keywords such as *formal methods, industrial automation, dependability,* and *model-based engineering* (a typical architecture for IAS) yielded more than 2000 works. This initial list was pruned to about 400 after using the phases of the SDLC and industrial automation standards such as IEC 61499 and IEC 61131-3 as additional keywords.[1] We further reduced the number of studies to 123 by focusing on offline formal approaches and also by studying how closely each work related to IAS by manually reading the titles, abstracts, and the conclusions sections of the articles. This survey is designed for industry practitioners who can evaluate available approaches for use in their contexts, as well as for researchers in industrial automation systems, dependability analysis, or formal methods, who would like to explore the intersection of these fields.

In Sections 2–4, we use the SDLC phases to categorize surveyed works, although some approaches span multiple phases. Focusing on individual phases allows us to see how well existing approaches address the key dependability-related challenges in each phase, which leads to a more thorough analysis of the state of the art and future directions. Section 3 combines both the design and implementation phases as most formal approaches for assuring dependability in one phase were found to be applicable in the other. In Section 5, we integrate the conclusions drawn within Sections 2–4 and provide a discussion around overall trends, research gaps, and future directions. Our findings include identifying SDLC phases that have found successful use for formal methods, the imbalance between techniques borrowed from other domains and those developed or customized specifically for IAS, and the challenges in making formal methods intrinsic to the IAS SDLC. These observations lead to identifying a number of promising research directions to accelerate the adoption of formal methods in IAS.

### A. Key Definitions
A lack of standardization around dependability in IAS means that several terms can have ambiguous meanings, depending on where we look. Hence, we precisely define some key terms used in the context of this paper.

The term *specification* refers to either *requirements specification* and/or *system specification*. The former is used in Section 2, whereas the latter is used in Sections 3–4. *Formal* refers to methods, models, and algorithms that have precise syntax and semantics, allowing unambiguous specifications and interpretations. Temporal logics are well-known formalisms to specify requirements. *Informal* refers to techniques, tools, and methods where specifications and interpretations can change subjectively between observers. For instance, requirements written in natural language are inherently informal. *Semiformal* tools and approaches provide some formalism, but not enough to allow a completely objective interpretation. Unified Modeling Language (UML), for instance, provides a formal structure for the formation of requirement specifications, but terms used in a UML diagram may have different interpretations.

In the literature, the terms *verification* and *validation* are frequently used interchangeably. We follow their distinct definitions from the IEEE Standard Glossary of Software Engineering Terminology [16]. Verification is the process of determining if the artifacts produced in the current phase of the SDLC fulfill the requirements established during a previous phase. In other words, verification checks for consistency or compliance between a more detailed *implementation* and a more abstract *specification*. Validation, on the other hand, checks if specifications accurately capture customer needs.

## 2. Requirements Engineering

Requirements engineering is the first phase in system development and involves the elicitation, analysis, specification, and validation of requirements [17]. Requirements express the needs and constraints on a system, and in IAS, requirements can correspond to the software, hardware, and/or the physical processes being controlled. In addition to providing a precise understanding of the system to be built, requirements engineering has a far-reaching impact on subsequent phases in system development via traceability of requirements to system artefacts, and change management [18]. Most current requirements engineering processes are informal or semiformal at best. For instance, requirements elicitation, where the aim is to capture as many requirements from stakeholders, is largely an informal process. Most formal approaches for requirements engineering have focused on requirements specification and requirements analysis [19].

### A. Requirements Specification

Requirements specification involves creating a structured system requirements specification (SRS) document that can be systematically reviewed and evaluated. SRS allows us to analyze requirements and to also estimate costs and risks. While requirements are often written informally, several *requirements specification languages* (RSLs) allow semiformal or even formal specification of certain classes of requirements.

In the realm of semi and fully formal requirements specification, UML and use cases and their variants are by far the most common RSLs. SysML is another widely accepted semiformal RSL in automation systems [20] and can capture functional and safety requirements. Other RSLs include SWSpec based on Petri nets for formally specifying requirements in service workflow environments [21].

Adequate user guidance can make the writing of formal requirements easier. Often this is done by providing templates or patterns for writing requirements. Such patterns for *possibility* or *fairness* are proposed in [22], which were later extended into a formal methodology for specifying real-time communication requirements for industrial energy distribution systems [23]. Template-based conversion of informal requirements to semiformal boilerplates and then to formal patterns was also proposed in the CESAR project [17].

Some approaches extract formal requirements from more informal sources automatically. In [18], semiformal engineering design process enables the conversion of computer aided design (CAD) documents to formal documents. An approach to formalizing requirements in natural language to formal specifications [24] uses lexical analysis to create a structure diagram for requirements. A language in which test information is extracted through the various stages of system development starting from the requirement stage is presented in [25].

### B. Handling Nonfunctional Requirements

A dependable system must satisfy NFRs, such as reliability, availability, safety, and security [1]. In the requirements engineering subphase, most existing works cover only safety. In [26], *safety* requirements over individual components and compositions are specified and verified. SysML is used to capture safety requirements in a tree structure containing relationships between requirements and subrequirements. This approach also provides support for traceability by allowing individual requirements to be linked to components that implement them. When requirements are refined, subrequirements are automatically reallocated to subcomponents and their combined behaviors are checked. The SAPIS tool allows formal specification of safety requirements according to the CENELEC standard and is based on previous work on safety [27]. A classification of safety requirements for industrial automation appears in [28]. Classifications include *static* requirements that demand global satisfaction of a property or *dynamic* requirements that must be true in certain states and false in others. These requirements can be easily translated to temporal logic formulas using templates. In [29], graphical formalisms traditionally used for modeling fault trees in safety analysis were used for modeling security attacks.

TABLE I. COMPARISON OF WORKS SURVEYED FOR REQUIREMENTS ENGINEERING

| Technique | Sub-phase | NFR type | Formal? | Tools | Industry |
|---|---|---|---|---|---|
| [20, 26, 27] | S | safety | Semi | ✓ | ✓ |
| [21] | S | - | Fully | ✓ | - |
| [17, 22] | S | general | Semi | ✓ | ✓ |
| [18, 24] | S | - | Semi | ✗ | ✗ |
| [29] | S,A | safety security | Semi | ✗ | ✗ |
| [25] | S,A | - | Semi | ✗ | ✗ |
| [20, 31] | A | traceability | Fully | ✓ | ✗ |

### C. Requirements Management and Analysis

Requirements management is concerned with maintaining requirements in a usable manner throughout the SDLC. One problem is storing requirements and associated artefacts such as test-cases and ensuring that they are reusable (concretized) as we go down the SDLC. In [25], a model-based methodology called test requirements model is used to transfer tests between various stages of the life cycle. It follows the behavior model for tests standardized by a consortium of standardization committees including IEEE.

Continuous validation and analysis are required as requirements get refined during the development of the system. Requirements traceability is useful as it allows linking requirements as they evolve [30]. Formal methods can help find consistencies between requirements. In [31], a trace creation and recovery approach using context analysis enables a model-based approach to validate requirements using timed state machines, provide feature-oriented requirements validation, and generate runtime observers for requirements. Requirements management requires managing changes in requirements. An approach based on SysML to depict and analyze change influences in industrial automation systems is presented in [20]. The focus is on modeling the impact of requirement changes on the mechanical, hardware, and software aspects of an IAS.

Management and analysis require maintaining the knowledge within requirements, which is often done using ontologies to store interrelationships between concepts. Formal methods can help maintain a structured ontology, interlinking requirements, and persistently maintaining these links [32].

Table I visually compares the surveyed works by the requirements engineering subphase they apply to (S—specification, A—analysis), the type of requirements (indicates functional requirements only), and whether each formalism is fully or semiformal. Additionally, we also show if there are available tools support and evidence of industry use for every technique. Overall, requirements specification and requirements analysis, are well covered by existing works. However, the lack of formalized requirements management frameworks is a significant gap in current research. Also, there is no existing framework covering a wide range of NFRs, which indicates that existing works may find only niche use. A further discussion of these results appears in Section 5.

## 3. Design and Implementation

During the design phase, the solution space described by requirements is increasingly constrained. During implementation, a finalized design is extended to build the system.

### A. Design

The design phase can be broken down into two broad categories: *high-level* design, followed by *low-level* design [17]. The key concern in high-level design is the creation of a system *architecture*. The system architecture captures an initial, abstract layout of the main subsystems or parts of a system via a primary separation strategy, and also contains architectural tactics to deal with primary quality or nonfunctional attributes of the system such as dependability. Existing standards such as IEC 61499 and IEC 61131-3 provide robust software architectures for IAS. A few formalized extensions or alternatives are reusable automation components [33], intelligent mechatronic components, and automation objects [34]. These architectures provide better formalization, increased flexibility, and the ability to design software in a more hierarchical manner. Model-driven design is a key primary separation strategy used in IAS architectures, where a system is broken down into a *controller* constituting the software and PLC hardware and a *plant* representing the physical processed being controlled.

In low-level design, an increasingly detailed layout of the system is built. Here, visual domain-specific languages (VDSLs) are quite common. Several works provide some kind of formalized modeling support. VDSLs in [35] come complete with syntax and behavioral semantics for converting designs to Petri nets while preserving timing or safety-related behaviors leading to early verification of designs. Early stage automata-based VDSLs presented in [36] also allow the modeling of safety but do not provide sufficient test data to back the theoretical basis. Continuous function charts [37] is a VDSL based on Statecharts that allows hierarchical designs of discrete-continuous embedded systems. It allows both control and data flow to be explicitly specified. These can potentially be used in IAS design too as models can be automatically translated into code that is amenable to coverage analysis.

Standards such as IEC 61499 and IEC 61131-3 are the most popular design languages for IAS. Both provide VDSL-like features for creating component interfaces and networks with an ability to embed code into components during implementation. A plethora of works exists in the formalization of these standards, such as formalizing the syntax and semantics of IEC 61499 presented in [38]. IEC 61499 suffered ambiguities in its execution semantics resulting in several works that formalize the execution model differently [39]. Subsequently, several further works focus on transforming IEC 61499 programs from one execution model to another [40]. A few works extend the semiformal SysML industry standard for modeling, such as the Manufacturing Execution Systems Modeling Language MES-ML [41] and extensions of SysML for modeling activities to enable safety analysis [42]. A comprehensive perspective on the whole analysis and design process during system development is presented. Works focusing on interoperability include [43] for re-engineering IEC 61131-3 programs into other paradigms such as IEC 61499 programs.

UML is used in several domains and we find several works on formalizing and adapting UML for designing IAS [44]. UML enables modeling of control logic [45], often leading to automatic code generation of standard PLC languages such as function blocks [46]. In [47], the IEC 61499 standard is extended to include UML-Statecharts features to reduce software complexity while ensuring automatic generation of standard-compliant code.

Some works use formal languages or a combination of formal languages and industry standards for specifying system designs. A specification language for control programs based on linear temporal logic and structured text called ST-LTL is presented in [48], to make it easier for control engineers familiar with structured text to formally specify their designs. Similarly, a B language based design specification can allow formally proving that a design provides completeness, consistency, precision, and correctness guarantees [49]. Other languages combine several formal languages, such as Petri nets and Object-Z [50], Specification and Description Language, and languages underpinning formal tools such as NuSMV and SIPN.

A large body of work exists in IAS design using Petri nets and their extensions. Petri net extensions include Hierarchical Colored Petri nets, Timed Petri nets, and High-level Petri nets. Some works combine Petri nets with other techniques such as supervisory control approaches [51] for design-level analysis, validation, and simulation of IAS. Several formal frameworks are based on Petri nets, such as timed net/condition event systems [52] for modeling PLC behavior, or signal net systems for modeling and verification of distributed control systems. Some drawbacks of these initial works, such as a lack of formalized semantics, have been addressed in later works [53]. Some works extend existing implementation approaches using Petri nets, such as [54] where the GHENeSYS modeling environment is extended by adding a function of process actions association with places and transitions. Some works study Petri nets based modeling patterns [55] for enabling model to code generations.

Plant modeling is an integral part of model-driven design. Current works using formal or semiformal frameworks for this purpose include using SysML variants to model changes in mechatronic production systems [20], closed-loop modeling of plant and controller using net condition/event systems (NCES) leading to formal verification [56] or code generation. In [57], an approach to synthesize discrete-state plant models from behavior traces and temporal properties is proposed. The problem is posed as a Boolean satisfiability problem (SAT) and is solved by running a SAT solver. The generated models are intended to be applied in closed-loop model checking.

TABLE II. COMPARISON OF DESIGN AND IMPLEMENTATION WORKS

| Technique | Standard | Model | Tools | Industry |
|---|---|---|---|---|
| [34] | IEC61499 | NCES | ✗ | ✗ |
| [35, 53, 55] | IEC61131-3 IEC61499 | Petri nets | ✓ | ✓ |
| [36, 37] | - | Automata | ✗ | ✗ |
| [38, 40] | IEC61499 | Algebraic | N/A | ✗ |
| [43] | IEC61131-3 IEC61499 | Ontology | ✗ | ✗ |
| [44–46] | IEC61131-3 | UML | ✓ | ✓ |
| [47] | IEC61131-3 | Statecharts | ✓ | ✓ |
| [48–50] | IEC61131-3 | various | ✓ | ✗ |
| [20, 56] | - | various | ✓ | ✓ |
| [59, 60] | IEC61131-3 | Petri nets | ✓ | ✗ |

**B. Implementation**

A small number of formal methods directly target IAS implementation. Several formal *controller synthesis* approaches have been proposed from formalized system models [58] and differ mainly in the kinds of formalized system models they require as inputs. Some other works convert formal specifications to standard-compliant code [59], [60], differing again in input specification and output language types. The method presented in [61] uses Petri net models for controller synthesis in commercial PLCs using IEC 61131-3 function blocks enhanced with object-oriented programming techniques to achieve event driven semantics for the target PLC hardware.

Some approaches infer executable models of off-the-shelf components with black-box interfaces through black-box testing data. The technique presented in [62] infers the behavior of components as IEC 61499 programs, but cannot guarantee completeness due to nonexhaustive exploration. In [63], an automated people mover system is developed and tested using both verification and simulation. Formal verification was used to check only for deadlocks, whereas timed models in UPPAAL were used for simulation purposes. In [64], implementation artefacts are specified using formal languages such as variants of process algebra.

Table II provides a summary of the most works surveyed in this section. Each technique targets a specific standard and relies on an underlying formal model. Petri nets are the most commonly used formal model. Some works have tool-support available and only a few have found use in industry. Very few works span both IEC 61131-3 and IEC 61499 standards and subsequently do not exploit the highly structured nature of these standards. Semiformal approaches, such as those based on UML or Statecharts have found more use in the industry due to better usability.

A deeper discussion of these results appears later in Section 5.

## 4. Testing

The testing phase checks if a system meets its intended requirements. Conventional testing involves stimulating the implemented system through varying inputs and observing its outputs and other qualities (such as timing). However, as exhaustive testing is very expensive, a significant part of testing is carried out through *simulation*. Here, test vectors are used to stimulate and verify the behavior of a system *model*, but not the implemented system. Simulation is usually faster than conventional testing but is more useful where abstract simulation results are guaranteed over the actual implementation. The third category of testing involves *formal verification* techniques where formal methods are used to prove the correctness of a formal model or specification of a system. Formal verification techniques can guarantee correctness of the formal specification due to full coverage, but are often limited to small to medium system models. Formal methods, which have traditionally found use in formal verification, also hold promise in accelerating conventional testing and simulation [65]. For instance, formal methods can be used to select right stimuli for conventional testing and simulation for better coverage and reducing testing times. However, a large gap exists between the potential of formal methods in testing, as evidenced by research, and their actual use in industry; in fact, only a few solutions find direct industry use [66].

**A. Formal Methods in Conventional Testing**

Conventional testing involves stimulating a part or whole of a system implementation using a carefully chosen, automatically generated, or randomly selected sequence of stimuli or system inputs, called a *test case*. Formal techniques such as symbolic execution, model-based testing, combinatorial, random, and search based testing [66] can help automatically generate test cases. Other techniques can help with executing tests and providing tool support [67].

In model-based testing, incomplete system models are used to generate test cases to be tested on implemented control software. In [68], a system software specification in sequential function charts is translated into labeled transition systems and then traversed by a custom test tool to derive test cases. In [69], a hybrid system model is converted into a target discrete diagnostic model in Ludia, a language for modeling complex systems for fault diagnosis. This technique lacks a proof of correctness. At the component level, function blocks in both IEC 61131-3 and IEC 61499 provide modules that can be seen as black boxes amenable for test-case generation. In [70], a symbolic execution based approach is presented where an intermediate model from Arcade.PLC [71] is used to autogenerate test cases for IEC 61131-3 programs. This paper can test for unreachable code, a rarely encountered feature in other works. In [72], IEC 61131-3 function blocks are converted into timed automata models. Tailored unit tests and coverage requirements are generated in the UPPAAL modeling environment. A framework for verifying IEC 61131-3 instruction list programs is presented in [73]. It proposes a formal semantics for a significant fragment of the instruction list language, and a direct mapping of the semantics into a model checking tool. A requirements-based test-generation and tracking simulation based execution technique for systems built using IEC 61131-3 appears in [74].

Some works, such as [75], target test-case generation for complete systems. Here, a system under test (SuT) is described using executable networks of either IEC 61131-3 or IEC 61499 function blocks. A UML description of the system is used to generate tests in UML, and the test-case generation strategy traverses branches in the networks to ensure that each branch is visited once. Then, through model-to-model transformation or model viewing, the SuT, as well as generated test-cases, are transformed into IEC 61499 or IEC 61131-3 systems. A test-case generation strategy presented in [76] extracts test cases from IEC 61499 programs specified as state and activity diagrams using the round-trip path coverage strategy, where every defined sequence of transitions that begin and end in the same state is taken. Future directions in system testing include developing formal testing platforms to decide which components to test together, how they should be tested, and when to stop testing.

There are a number of tools for test-case generation and execution. Some tools specialize in generating and executing test cases for specific platforms, such as testing the timing of controller area network (CAN) bus based control applications [77], and the safety of software deployed on FPGAs [78].

### B. Simulation

Simulation involves using test data to stimulate a system model, not an actual implementation, in an attempt to expedite testing. Some approaches, such as [79], model communication tasks within an IAS using a semiformal language such as SysML out of which timed Petri net models can be extracted and simulated to estimate the performance of a distributed system. In such approaches, networks are replaced by models, and simulation parameters such as packet drops can be easily controlled.

Other simulation frameworks provide testing systems for NFRs and attributes. In [80], a random testing framework for safety-critical embedded systems is presented where constrained random testing enables larger coverage through random input variations, random fault injections, and automatic output comparisons. Functional safety standards, such as EC 60812, 61580, MIL-STD-1629A, etc., apply to many IAS [81]. Current approaches use safety PLCs for mission critical applications and formal failure mode and effects analysis approaches to analyze system hardware, but no existing formal technique can assess complete systems for functional safety. The need for a theoretical basis for formally modeling and assessing IAS for security requirements such as availability, integrity, confidentiality, graceful degradation, and detection is highlighted in [82]. The authors note the high modeling overheads that prevent the widespread use of formal modeling for this purpose.

### C. Formal Verification

Existing surveys such as [12] and [13] look at the use of specific formal verification techniques such as model checking in industrial automation systems. Distinctions between formal verification techniques arise from differences in system and requirements modeling, algorithms used, user effort required, and outputs produced. For instance, Xia *et al.* [83] propose an automatic verification tool for PLC systems, which includes formal modeling, static syntax checking, code generation and optimization, and visual representation of counterexamples.

Some works extract formal models of systems or requirements automatically. A syntactic and semantic analysis for IEC 61499 systems appears in [43]. A semantic checker carries out rule-based checking of software modules, constrained via a hard coding of correctness criteria to reduce user effort. In [84], timed models of plants and closed-loop automation systems are built as constrained timed discrete event systems and timed controller models, respectively. These implementation-independent models are subsequently transformed into timed automata for verifying timing requirements such as urgency. Other automatic modeling approaches, such as [85], target the conversion of software modules written in different languages to timed automata. A static code analysis tool is presented in [86] to parse and convert IEC 61131-3 control applications into control flow graphs. This model can then be assessed for type constraints, reachability, and liveness using a hybrid algorithm based on abstract interpretation and data flow analysis. A good summary of IEC 61499 modeling and verification can be found in [9]. A semantic characterization of PLC programs based on extended λ-calculus appears in [87], leading toward theorem proving and model checking.

Some works deal with graphical, intuitive, and more user-friendly front-end specification languages that are aimed at making formal methods more accessible. In [88], requirements specified using symbolic timing diagrams and safety-oriented technical language are translated to temporal logic leading to a model checking of closed-loop system models. Other works include specification patterns for the nuclear automation sector [89] and user-friendly visualization and handling of counterexamples generated by model checkers [90].

Formal verification algorithms face the well-known state explosion problem, where system model sizes explode when component compositions or data variables are taken into account. Typically, these details are either abstracted out or verification is focused on individual components to verify requirements. Several modeling languages such synchronous/asynchronous composition of finite-state machines and automata, networks of timed automata [91], and Petri nets [92] are being used in IAS to address state explosion.

Formal plant modeling has been a recent focus area for several works. It helps reduce the size of the formal model to be verified through closing an IAS system through a constrained plant model [93]. A good overview of plant modeling in formal verification can be found in [94]. In [95], a workflow to specify safety-critical systems, plants, and requirements as sequential function charts is proposed. These models are automatically translated into controller code, plant models (via library matching of modules), and requirements in temporal logic. These can then be simulated and verified using UPPAAL. A framework for modeling complete IAS, including plants as hierarchical and compositional NCES, presented in [96] provides support for model editing, visualization, and verification.

TABLE III. COMPARISON OF SURVEYED WORKS FOR TESTING

| Technique(s) | Type of Testing | Formalism | Standard | Formal? | Tools | Industry use |
|---|---|---|---|---|---|---|
| [67–69] | Conventional (Test Case Generation) | Transition systems | IEC 61131-3 | Semi formal | ✓ | ✗ |
| [72, 73] | Conventional (Test Case Generation) | Automata | IEC 61131-3 | Semi formal | ✓ | ✗ |
| [74, 75] | Conventional (Test Case Generation) | Tracability | IEC 61131-3 | Formal | ✓ | ✓ |
| [80] | Simulation | SysML | | ✓ | | ✗ |
| [85, 86] | Formal Verification (model checking) | Timed Automata | IEC61131-3, IEC61499 | Formal | ✓ | ✓ |
| [88] | Formal Verification (theorem proving) | Lambda calculus | IEC 61131-3 | Formal | ✓ | ✗ |
| [89, 91] | Formal Verification (model checking) | NCES | IEC 61499 | Formal | ✓ | ✗ |
| [94, 96, 97] | Plant modelling | Mixed | IEC61131-3, IEC61499 | Formal | ✓ | ✗ |
| [106, 121] | Formal Verification (model checking) | SMV | IEC61131-3, IEC61499 | Formal | ✓ | ✓ |
| [107–109] | Formal Verification (model checking) | Timed Automata | IEC61499, IEC61131-3 | Formal | ✓ | ✗ |

Some works target the formalizing of IAS standards. In [97], the Coq theorem prover is used to formalize the semantics of the Instruction List (IL) and Sequential Function Charts (SFC) languages from IEC 61131-3. Such frameworks enable automatic generation of code from graphical system specifications.

Several works transform system artifacts into format accepted by formal tools. In [98], IEC 61499 systems are modeled as finite-state machines, contracts, or modules in synchronous languages, subsequently allowing the use of model checkers. A few IAS-only model checkers exist, such as ViVe/SESA, VEDA [99], and Arcade.PLC [71]. Some works focus on converting deployable code (or block-based application design) into languages accepted by model checkers such as symbolic model checking (SMV) and UPPAAL, to enable automatic verification. The PLCverif [100] tool converts IEC 61131-3 programs to target model checking languages (NuSMV/NuXmv, UPPAAL). PLCverif has been subsequently extended to provide support for all five programming languages of the IEC 61131-3 standard [101], [102]. A transformation of IEC 61131-3 programs into PVS specifications to allow theorem proving is presented in [103]. This was done to mandated requirement of applying formal verification to such safety systems. In [104], rule-based model transformations help convert IEC 61499 programs into SMV models for NuSMV/nuXmv-based model checking.

Many algorithms use the UPPAAL model checker to verify timeliness. In [105], IEC 61499 systems are transformed into timed automata using set translation rules for verifying timing requirements. However, hardware configurations that affect a system's timing characteristics are not considered. Other works include using hybrid automata to analyze the worst case execution times in component networks [106], and timed automata to elevate existing IEC 61131-3 systems for compliance with newer safety standards such as IEC 61508 [107]. Such approaches feature rule-based model transformations. In [108], formal dependability analysis is carried out by manually modeling input/output modules both qualitatively as point systems and quantitatively using probabilistic to compute metrics such as system reliability. In [109], client-server networked automation systems are evaluated by using timed event graph models that are analyzed using deterministic and probabilistic analyses. Mazzolini et al. [110] show an early stage verification strategy for IAS using model checking algorithms and model coverage.

Some approaches focus on specific subproblems such as scheduling of flexible manufacturing systems by using formal analysis to assign schedules to a system divided into subnets [111]. Some works, such as [112], formalize the execution semantics of flexible manufacturing systems modeled using stateflow diagrams and can reason about undesirable execution sequences that ambiguities in the specification language can induce. The formal analysis of a listen-before-talk protocol for interacting components over congested wireless networks is presented in [113]. Assuring network performance is a significant problem in general. In [114], several model checkers are used for the security analysis of industrial wireless protocols.

A few approaches look at formally verifying NFRs. Online reconfiguration of IAS is an important topic addressed in works such as [115]. The work in [116] looks at formally specifying buffered sequential execution model of the Fuber IEC61499 runtime. It also presents extended finite automata models that are suitable for formal verification of the proposed execution semantics. In [117], industrial controllers, modeled as finite-state machines can be verified for fault tolerance using SMV. Techniques such as [118] explore the use of formal methods to verify the security of networks used in industrial settings. In this particular work, $\pi$ calculus is extended with security features allowing system and requirements modeling, which enables automatic verification of vulnerabilities. A fault-tolerance analysis of control algorithms using signal interpreted Petri nets from [119] uses formal modeling and SMV to check for determinism.

Table III shows a comparison of some of the works covered in this section based on the types of testing, formalisms used, and the target IAS standards. We also list whether a technique is semi or fully formal, has tool support available and if it is used in industry. Semiformal techniques do not directly use mathematical formal models, for example, most transition systems such as Petri nets are classified as semiformal due to lack of formal analysis of transition rules. Most techniques have robust tool support, unlike techniques that target earlier SDLC phases. We also note that several works propose transforming one model type to another to enable the use of additional analysis tools. Often, these approaches lack rigor in the transformation process. The most successful formal methods target conventional testing and aim to make it easier, more comprehensive, and/or faster. A more detailed analysis of our findings appears in Section 5.

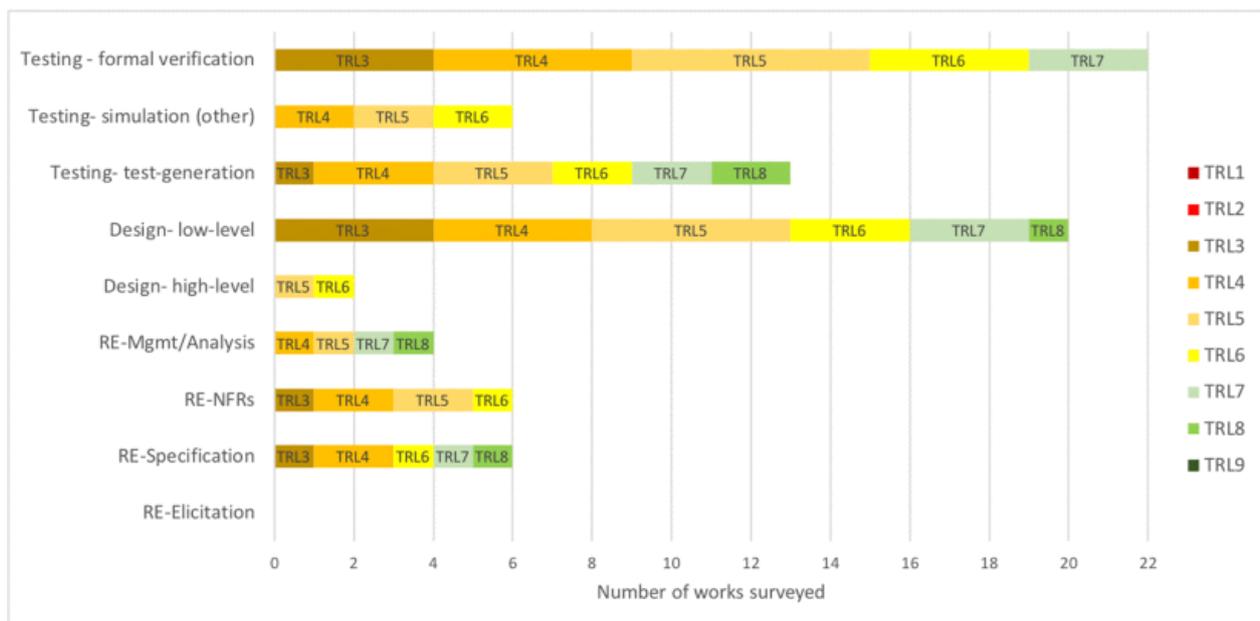

Fig. 1. Proliferation and technology readiness of surveyed works, categorized by SDLC subphases.

## 5. Discussion

This survey reveals some key insights into the current use of formal methods for dependability in IAS. Fig. 1 shows the number of formal methods surveyed for each subphase of the SDLC. These numbers provide a clear picture of the subphases, such as design and testing, that have found more use for formal methods. Fig. 1 also shows the technology readiness level [120] (TRL) for each approach we studied. We use the TRL definitions [121] adopted by the European Commission since 2014 to derive and classify the articles into different TRL levels. TRL provides a scale to assess the maturity of a technology. Lower levels, such as TRL1–3 indicate that a work is more conceptual, whereas higher levels TRL8–9 indicate robust, industry-ready technologies that have sufficient process and tool support available. Overall, most surveyed works were in the TRL3–5 range, indicating the relatively low maturity of formal methods in IAS. This also corroborates with the split between early-stage works and established works. We cite 47 (38% or all citations) early-stage papers that are mostly conference papers proposing techniques at lower TRLs. The remaining 62% are established works at higher TRLs and published mostly in established journals, as well as a few high impact conferences, white papers, and patents. In the following paragraphs, we unpack the information contained within Fig. 1 and derive critical information about the state of the art in the use of formal methods for each SDLC phase.

Requirements engineering, the first phase of the SDLC, involves eliciting, organizing, specifying, analyzing, and managing requirements. Some aspects of this phase, such as elicitation, are inherently informal and do not utilize formal methods at present. However, there is immense value in the early use of formal methods for requirements specification, management, and analysis. The earlier we can unambiguously capture the knowledge base a system is built on, the easier and cheaper it becomes to build the system. Several existing formal and semi-formal methods provide sufficient support to model requirements, the knowledge these requirements is built on (ontologies), and for automatic or guided analysis. However, their use is sporadic due to the high learning curves and expert skill sets needed. While the lack of tool-support is an important issue, standardization of the use of formal methods for various specification, analysis, and management of requirements can go a long way in addressing this problem. Both tool-support and standardization can help build a critical mass of repeatable and feasible tools, processes and skills in the domain. Standardization will involve formalizing common domain-specific knowledge in IAS, as well as lay out a roadmap for incorporating new formal methods into the requirements engineering phase. We also find inadequate formal support for NFRs. Again, the application of standardized formal models, templates, and analysis methods for IAS-specific NFRs, such as safety, dependability, and timing, can help with wider adoption and use of formal methods. The most robust technology in this area is the semiformal SysML, which is an industry-wide standard for modeling behavioral and some safety requirements. Other technologies lie more in the proof-of-concept or early validation stages and require more experimentation with real-life systems.

In the design phase, primary functional requirements and primary NFRs are converted into a structured solution space that is then concretized in the implementation phase. This phase is a crowded field, with most formal methods providing support for modeling techniques. However, only a few provide a seamless transition from the requirements engineering phase, making them more useful than other, more disjointed or stand-alone modeling techniques. Similarly, modeling techniques are more useful when they can also be used in the implementation phase such as several works that support automatic code generation. As the system gradually crystallizes during these phases, frameworks to formally trace and map requirements through these phases can be extremely useful for ensuring consistency between subsequent system models or implementations. Given that, currently, requirements, design, and implementation aspects are typically specified using different formal tools and techniques, integrating these into the SDLC is both an existing problem and an exciting direction for future research. It is infeasible to expect a single modeling framework for all these stages. Hence, we hypothesize that it is more important to focus on formal model transformations, such as in model-driven design to bring in much needed compatibility between different frameworks. Mature design and implementation standards such as IEC 61499 in IAS provide a structured foundation on which to build such model transformations. Further formalization of these standards is, therefore, another interesting direction to improve the uptake of formal methods. Formal methods targeting requirements standards such as IEC 61508 will also help improve support for building dependable IAS. From a technology readiness perspective, VDSLs based on semiformal SysML and UML are used widely in industry in the design phase. Some formal methods such as Petri nets and those involving plant modeling have also grown in popularity and have found wider use in both design and implementation.

The testing phase of the SDLC has found the most use for formal methods. Conventional testing is aided by formal methods through comprehensive test-case generation, selection, and execution. For test generation, we find that most frameworks are semiformal, and there is need to formalize them more to derive coverage guarantees. A few approaches help with testing implementations against NFRs such as fault tolerance and functional safety. Considerably more work is needed to test for other important NFRs, such as reliability, availability, and most importantly, security. Security is a growing concern in this area since IAS have moved on from being contained within factory walls to being large, cloud-based solutions. In simulation-based testing, some formal tools and frameworks such as UPPAAL has found more popularity in timing-related verification of system models in IAS. Otherwise, the landscape remains fairly fragmented as often models and tools are selected for very specific problems and hence are not generalizable. In a similar vein, we also found several formal verification-based testing tools, differing mostly in requirement types and models, and the format of systems supported. A bulk of the work in this area has focused on plant modeling and verifying both plants and controllers together, highlighting the domain's focus on formally capturing the physical processes controlled by IAS. In general, though, formal verification based testing techniques have found limited industrial adoption due to well-known limitations such as state explosion and required user-expertise. Fig. 1 clearly shows that most formal verification techniques in the testing phase are at low TRL levels.

## 6. Conclusion

This paper presents a survey of static, formal techniques for building dependable IAS. Static approaches are more useful during the earlier stages of the SDLC; bugs and inconsistencies are cheaper to find and correct during these stages. Online approaches are equally useful but are used to analyze system executions and have been surveyed elsewhere. Using the phases of the SDLC for categorizing the approaches we surveyed, we find that most works apply to the design and testing phases. On the other hand, other SDLC phases such as requirements engineering contain wide gaps and provide interesting opportunities for future investigation. A closer look at the surveyed works reveals that most works do not seamlessly integrate with current industry practice, and have not matured sufficiently for use in system development. However, the picture is not universally bleak. IAS standards for requirements, design, and implementation, provide inherent support for formal approaches through their structure. Additionally, semiformal approaches provide a balance between rigor and usability, and some of these have matured enough for industry-driven use during the various SDLC phases.

Going forward, there is a need to carry out substantial *evaluative* research, where the effectiveness of available formal methods must be tested empirically in industrial scenarios. This is contrary to the more prevalent *propositional* approach, where new models, algorithms, and formalisms are being proposed rapidly, resulting in a sparse landscape with little to no industry adoption. The evaluative approach will help ensure that more robust formal methods for the IAS domain incorporate domain-specific knowledge and context, allowing for more repeatable use.

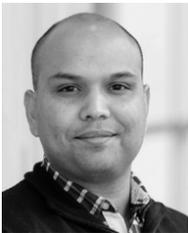

**Roopak Sinha** (S'03–M'13) received the Ph.D. degree in electrical and electronic engineering from The University of Auckland, Auckland, New Zealand, in 2009.

He is currently a Senior Lecturer with the School of Engineering, Computer and Mathematical Sciences, The Auckland University of Technology. He has previously held academic positions with The University of Auckland, and INRIA, France. His research interests include "Systematic, Standards-First Design of Complex, Next-Generation Embedded Software Âİ applied to domains such as Internet of Things, edge computing, cyber-physical systems, home and industrial automation, and intelligent transportation systems.

Dr. Sinha has served on several IEEE/IEC standardization projects, was an invite coeditor of a Special Section on "Dependability in Industrial Informatics" of the IEEE Transactions on Industrial Informatics, and works with several New Zealand companies to systematically reduce standards-compliance costs in IoT/embedded products.

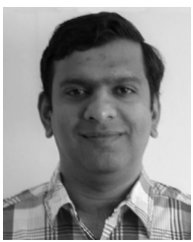

**Sandeep Patil** (S'11–M'19) received the Ph.D. degree in computer science engineering from Luleå University of Technology, Luleå, Sweden, in 2018.

He is currently a Postdoctoral Researcher with the Dependable Communication and Computation Systems Group, Luleå University of Technology, Luleå, Sweden. His research interests include programming distributed industrial automation software systems using IEC 61499 standard. He is an accomplished software engineering professional with more than 10 years of research and development experience in systems and application software, including four years at Motorola India Pvt., Ltd., India, as a Senior Software Engineer.

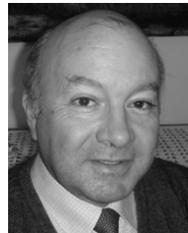

**Luis Gomes** (SM'06) received the Electrotech. Eng. degree from Technical University of Lisbon, Lisbon, Portugal, in 1981, and the Ph.D. degree in digital systems from NOVA University of Lisbon, Lisbon, Portugal, in 1997.

He is currently a Professor with the Electrical and Computer Engineering Department, NOVA School of Science and Technology, NOVA University of Lisbon, Portugal, and a Researcher with the Centre of Technology and Systems, UNINOVA Institute, Portugal. From 1984 to 1987, he was with EID, a Portuguese medium enterprise, in the area of electronic system design, in the R&D engineering department. His main interests include the usage of Petri nets and other models of concurrency, applied to reconfigurable and embedded systems codesign and cyber-physical systems. He has authored and coauthored of more than 200 papers and chapters published in journals, books, and conference proceedings and one book and has coedited three books.

Prof. Gomes was made Honorary Professor by Transilvania University of Brasov, Brasov, Romania, in 2007, as well as Honorary Professor of Óbuda University, Budapest, Hungary, in 2014. He was a recipient of the IEEE Industrial Electronics Society Anthony J Hornfeck Service Award in 2016.

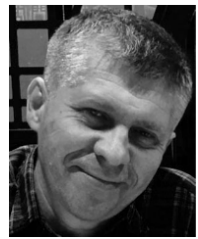

**Valeriy Vyatkin** (M'03–SM'04) received Ph.D. degree in applied computer science from Taganrog State University of Radio Engineering (TSURE), Russia, in 1992, the Dr.Eng. degree in electrical engineering from Nagoya Institute of Technology, Nagoya, Japan, in 1999, the Dr.Sc. (Eng.) degree in information and control systems from TSURE, in 1999, and the Habilitation degree from the Ministry of Science and Technology of Sachsen-Anhalt, Germany, in 2002.

He is on joint appointment as the Chair of Dependable Computations and Communications, Luleå University of Technology, Luleå, Sweden, and a Professor of information technology in automation, Aalto University, Aalto, Finland. He is also the Co-Director of the International Research Laboratory "Computer Technologies," ITMO University, Saint-Petersburg, Russia. Previously, he was a visiting scholar with Cambridge University, Cambridge, U.K., and had permanent appointments with the University of Auckland, New Zealand, Martin Luther University, Germany, as well as in Japan and Russia. His research interests include dependable distributed automation and industrial informatics; software engineering for industrial automation systems; artificial intelligence, distributed architectures and multiagent systems in various industries: smart grid, material handling, building management systems, datacentres and reconfigurable manufacturing.

Dr. Vyatkin was a recipient of the Andrew P. Sage Award for the best IEEE Transactions paper in 2012. He is the Chair of the IEEE Industrial Electronics Technical Committee on Industrial Informatics.